# Localized holes and delocalized electrons in photoexcited inorganic perovskites: Watching each atomic actor by picosecond X-ray absorption spectroscopy

Fabio G. Santomauro[1], Jakob Grilj[1], Lars Mewes[1], Georgian Nedelcu[2,3], Sergii Yakunin[2,3], Thomas Rossi[1], Gloria Capano[1], André Al Haddad[1], James Budarz[1], Dominik Kinschel[1], Dario S. Ferreira[4], Giacomo Rossi[1,*], Mario Gutierrez Tovar[1,†], Daniel Grolimund[4], Valerie Samson[4], Maarten Nachtegaal[4], Grigory Smolentsev[4], Maksym V. Kovalenko[2,3] and Majed Chergui[1]

[1] Laboratoire de Spectroscopie Ultrarapide, ISIC-FSB and Lausanne Centre for Ultrafast Science (LACUS), Ecole Polytechnique Fédérale de Lausanne, CH-1015 Lausanne, Switzerland

[2] Institute of Inorganic Chemistry, Department of Chemistry and Applied Biosciences, ETH Zürich, CH-8093 Zürich, Switzerland

[3] Laboratory for Thin Films and Photovoltaics, Empa − Swiss Federal Laboratories for Materials Science and Technology, CH-8600 Dübendorf, Switzerland

[4] Paul Scherrer Institut, CH-5232 Villigen, Switzerland

Majed.chergui@epfl.ch

We report on an element-selective study of the fate of charge carriers in photoexcited inorganic $CsPbBr_3$ and $CsPb(ClBr)_3$ perovskite nanocrystals (NCs) in toluene solutions using time-resolved X-ray absorption spectroscopy with 80 ps time resolution. Probing the Br K-edge, the Pb $L_3$-edge and the Cs $L_2$-edge, we find that holes in the valence band are localized at Br atoms, forming small polarons, while electrons appear as delocalized in the conduction band. No signature of either electronic or structural changes are observed at the Cs $L_2$-edge. The results at the Br and Pb edges suggest the existence of a weakly localized exciton, while the absence of signatures at the Cs edge indicates that the $Cs^+$ cation plays no role in the charge transport, at least beyond 80 ps. These results can explain the rather modest charge carrier mobilities in these materials.

---

[*] On leave from : Dipartimento di Fisica e Astronomia, Alma Mater Studiorum - Università di Bologna Viale Berti Pichat 6/2, 40127 Bologna, Italy

[†] On leave from : Departamento de Química Física, Facultad de Ciencias Ambientales y Bioquímica, y INAMOL, Campus Tecnolóogico de Toledo, Universidad de Castilla La Mancha (UCLM), Avenida Carlos III, S.N., 45071 Toledo, Spain.

# I. Introduction

Perovskites are a class of crystalline materials with the formula $ABX_3$, in which A and B are cations and X represents an anion. Lead-halide organic-inorganic perovskites (in which A is a monovalent organic cation) have recently emerged as highly promising optoelectronic materials for photovoltaics, photodetection, light-emitting diodes and laser devices.[1-8] Solar cells with lead-halide organic-inorganic perovskites (*e.g.* $CH_3NH_3PbI_3$), prepared by low-temperature solution-based methods, have recently achieved outstanding performances with photo-conversion yields exceeding 22%.[9, 10] This high conversion efficiency, together with the low processing temperature (150 °C), make them ideal candidates for the development of low-cost solar cells.

Despite the intense interest for organic-inorganic perovskites, a clear understanding of the fate of the photoinduced charge carriers remains limited. Lead-halide organic-inorganic perovskites exhibit a sharp absorption onset at the optical band edge, with a small extension of the Urbach tail (~15 meV).[11, 12] The photoluminescence (PL) spectrum is near the gap (i.e. weakly Stokes-shifted) and is homogeneously broadened by interactions with phonons, leading to a considerable intensity beyond the band edge.[13] Additionally, long carrier lifetimes and low non-radiative losses have been reported,[14, 15] pointing to a low density of traps (~$10^{10}$ cm$^{-3}$ in single crystals)[16-19] and leading to high PL quantum yields (>70%) for organic-inorganic perovskites.[15, 20] Nevertheless, the charge carrier mobilities in these perovskites are rather modest,[21] and in the case of the electrons, at least 1 order of magnitude lower than in materials such as silicon or gallium arsenide. Furthermore, as far as applications are concerned, understanding the recombination dynamics of photogenerated charges is important and in particular, which of the excitons or free carriers is the dominant recombination channel. In general, photovoltaic materials require efficient separation of photocarriers and the exciton binding energies (BE ≈ 10-70 meV), reported for organic-inorganic perovskites,[22-24] are on the same order of magnitude as the room temperature (RT) thermal energy, raising the question whether the excited states dissociate into free carriers or recombine radiatively. This fundamental question also relates to the issue whether transport of energy occurs by free carriers or by a bound exciton that later dissociates into free carriers at heterojunctions.[25-27]

PL studies have been performed to address these questions. In refs[28-31], the RT PL was attributed to exciton recombination but this was based on indirect evidence, while in ref. [22], it was attributed to free carrier recombination, arguing that excitons generated by low-density excitation are almost fully ionized at RT since the BE is comparable to the RT thermal energy.[11,2316] Recently, He *et al* suggested that the RT PL is due to weakly localized excitons, while they observed the free carriers emission at higher energies under high fluences.[32] They suggested that the weakly localized excitons are due to band tail states, presumably arising from the disorder introduced by the organic cation. These excitons can be either partly localized (one charge carrier is localized with another carrier bound to it by Coulomb attraction) or fully localized (both charge carriers are localized).

Organic-inorganic perovskites have raised concerns about their long-term thermal stability under the working conditions of solar cells.[33-35] Migration of the methyl ammonium cations ($CH_3NH_3^+$) is a factor of degradation,[36] while moisture can catalytically deprotonate them.[37] These last two problems are minimized by replacing the organic cation by an inorganic one, such as Caesium, as this leads to more stable materials under long term irradiation.[35, 38-42] Stoichiometrically, the inorganic $CsPbBr_3$ material consists of one $Cs^+$ and one $Pb^{2+}$ cation for three halides (e.g. $Cl^-$, $Br^-$ or $I^-$), as depicted in Figure S1, which form a cubic crystal structure for bulk samples (Figures S3 and S4). In the case of the mixed-halide $CsPb(ClBr)_3$ material, the composition is of about 2 $Br^-$ ions to one Cl$^-$. Inorganic perovskites are emerging as a very promising alternative to the organic-inorganic ones in solar energy conversion.[35, 40, 42] Furthermore, the ease of tuning their band gap properties by controlling the halide stoichiometry[38, 43, 44] holds great promise for successful use in tandem solar technologies.[35, 39]

In many respects, the physics of organic-inorganic perovskites and inorganic ones present several analogies. Photoemission and inverse photoemission studies have shown that they have identical electronic structures,[45] and this is supported by calculations of the electronic band structure of CsPbBr$_3$ and CsPbCl$_3$,[38, 45-47] which show similar trends as in the case of organic ones.[8, 48-52] In these systems, the top of the valence band (VB) is dominated by the halide p-orbitals (I 5p, Br 4p or Cl 3p), with a weak contribution of Pb 6s orbitals, while the bottom of the conduction band (CB) is dominated by the Pb 6p-orbitals, with a weak contribution of halogen p-orbitals.[38, 47] Finally, the orbitals of the cation (methylammonium or Cs) are either far below the maximum of the VB or far above the minimum of the CB. Another common aspect to both types of perovskites is their high PL quantum yield, which reaches 90% in the inorganic case[38] and 70% in the organic-inorganic case.[15, 20] Finally, from a dynamical point of view, transient absorption (TA) studies from the femtosecond to the nanosecond regime also point to a similar overall behavior of the charge carriers in organic-inorganic[53-55] and inorganic ones.[56, 57] Therefore, due to their greater stability, their simpler structure and ease of synthesis, inorganic perovskites are emerging as the next generation of materials for solar energy conversion.

However, and just as with organic-inorganic ones, the fate of the photoinduced charge carriers needs to be clarified. This requires tools that are site-sensitive and/or element-specific and can probe the system at RT. Electron paramagnetic resonance (EPR) has been used to determine charge localization in solar materials[58] but it requires low temperatures (LT). Shkrob et al.[59] studied by EPR LT (<200 K) $CH_3NH_3PbI_3$ and $CH_3NH_3PbBr_3$ samples irradiated at 355 nm and reported localization of holes at the organic cations and of the electrons at $Pb^{2+}$ centres, which formed clusters. This along with a change of color of the crystal surface during the measurements suggests that the samples were undergoing radiation damage. Aside from these considerations, EPR lacks the time resolution that would help identify the formation and evolution of charge carriers (e.g. trapping at defects or via coupling to phonons, etc.).

In recent years, time-resolved (from hundreds of femtosecond to tens of picosecond resolution) X-ray absorption spectroscopy (TR-XAS) has emerged as an ideal tool for probing charge transfer processes in a wide variety of molecular, biological and nano-systems.[60-62] In particular, we recently implemented it to probe the electron trapping at defects in RT bare and dye-sensitized anatase $TiO_2$ colloidal nanoparticles with picosecond[63] and femtosecond[64] time resolution. Photo induced $Ti^{3+}$ traps were identified with those formed by electron delivery via inter-gap excitation to the CB distinguishable from those delivered by interfacial electron injection.

In this communication, we investigated the fate of the charge carriers in photoexcited inorganic $CsPbBr_3$ and $CsPb(ClBr)_3$ perovskites by interrogating each atom of the material using time-resolved XAS with 80 ps resolution at the Br K-edge (near 13.474 keV), the Pb $L_3$-edge (near 13.035 keV) and the Cs $L_2$ edge (near 5.359 keV) perovskite nanocrystals (NCs).[38] $CsPb(ClBr)_3$ is a material with a band-gap energy and an exciton BE between those of the pure brominated and pure chlorinated Cs-based perovskites and we included it here in order to explore the role of halogen substitution on the fate of the charge carriers. Indeed, halogen substitution shifts in energy the VB more than the CB, as seen from calculations on organic-inorganic perovskites.[65] Our results show that while for both systems, the hole is localized at Br atoms in the VB, electrons remain delocalized in the CB and Cs atoms exhibit no response to photoexcitation.

## II. Experimental procedures

The synthesis and characterization (transmission electron microscopy, X-ray powder diffraction, UV-visible spectroscopy and XAS) of the perovskite NCs studied here are described in the § S1 and in Ref. [38]. The NCs involved in this work have an average size of ~12 nm (Figure S2) and were suspended in toluene. The solution was continuously refreshed in a free flowing liquid jet of 200 μm thickness, to ensure maximum sample stability during the laser pump/X-ray probe measurements. The time-resolved XAS experiments (more details in § S2) were carried out using the high repetition rate scheme at the microXAS and SuperXAS beamlines of the Swiss Light Source (SLS).[66] The pump laser ran at a repetition rate of 260 kHz delivering 10 ps-duration excitation pulses at 355 nm, which is above the band gap of the samples (Figure S5). The temporal width of the X-ray pulses is 80 ps at the SLS, which defines the time resolution of the experiment.

We recorded the UV-visible and XAS spectra before and after each measurement series (Figures S5 and S6), and they show that hardly no damage occurs to the sample. Finally, the pump fluence dependence of the transient XAS signal (Figure S7) exhibits a linear dependence up to at least 15 mJ/cm$^2$. Therefore, the latter pump fluence was used for the measurements.

## III. Results

Figure 1 shows the steady state Br K-edge absorption spectrum of CsPb(ClBr)$_3$ NCs and Figure S8a compares it with that of CsPbBr$_3$, detected in partial fluorescence yield (PFY) mode (see § S.2). These spectra are somewhat different when recorded in total fluorescence yield (TFY) mode (see § S.2), as can be seen in Figure S6. They resemble the spectrum reported for organic $CH_3NH_3PbBr_3$ perovskites.[67] Interestingly, the Br K-edge absorption spectrum of the $Br^-$ ions in the present samples is quite similar to that of the aqueous $Br^-$, also recorded in PFY mode.[68] Since $Br^-$ has a fully filled $4p^6$ valence orbital, the signal is entirely due to above-edge transitions into the ionization continuum with the X-ray absorption near-edge structure (XANES) modulations rapidly damping out.

Figure 1 also shows the transient Br K-edge spectrum (difference spectrum of the X-ray absorbance of the excited minus the unexcited sample) of $CsPb(ClBr)_3$ and $CsPbBr_3$, recorded 100 ps after laser excitation. The two transients exhibit the same profile, though with somewhat different amplitudes, showing an increased absorption below and above (around 13.485 keV) the edge and a decreased one at the edge (13.480 keV). At higher energies, clear modulations show up. As can be seen from Figure S9, the energy transients are identical at 100 ps and 1 ns time delay, to within error bars.

The time evolution of the transient X-ray signal was recorded at 13.480 keV and is shown in Figure 2 for both samples. To within error bars the time traces are quite similar, with a drop of intensity within the first 5 ns, followed by a long decay component stretching over 100 ns. However, at sub-ns times, differences show up with a slower decay for the Cl-free sample (Inset to Fig. 2). The similar kinetics of the signal (Figure S10) at the edge (13.480 keV) and above it (13.4863 keV) confirm that the spectral changes in figure 1 belong to the same transient species.

Figure 3 shows the steady state Pb $L_3$-edge spectrum of $CsPb(ClBr)_3$, and Figure S8b compares it with that of $CsPbBr_3$. The two spectra are identical, and are also quite similar to the Pb $L_3$ spectrum reported[67] for $CH_3NH_3PbBr_3$ nanoparticles. They are overall rather featureless, which is typical of Pb L-edge spectra (figure S11).[69, 70] Figure 3 also shows the Pb L$_3$-edge transients obtained for the two samples. The transients show some similar trends for both samples, bearing in mind the large error bars, due to higher noise: a decreased absorption right at the top of the edge (near 13.05 keV), while the signal appears to be positive above 13.065 keV. The signal also appears positive below the edge for the case of CsPbBr$_3$. Due to the low signal, we could not record time traces.

Finally, Figure S12 shows the Cs L$_2$-edge, which is characterized by a prominent "white line" at 5.363 keV, followed by a rather featureless above-edge region. Since Cs is in the +1 oxidation state, the white line edge is due to the 2p$_{1/2}$ to 6s transition. Under 355 nm excitation, no photo induced changes are observed at the Cs L$_2$-edge, as seen in Figure S12, contrary to the case of the Br and Pb edges. We therefore conclude that the Cs atoms are not affected by photoexcitation at times > 100 ps and we discuss this observation below.

## IV. Discussion:

In the following we discuss the nature of the photo-induced changes observed at each atomic edge. Generally speaking, oxidation state changes of an element show up as a shift of the edge position (i.e. the ionization potential for the atomic core orbital in question), which increases or decreases in energy depending, respectively, on whether the atom has been oxidized or reduced.[71] However, the origin of the edge shift contains other causes, e.g. the type of ligand and the bond distance between the atom of interest and its neighbors.[72-74] Photoinduced structural changes will therefore have a direct impact on the edge, even in the absence of an oxidation state change.[75, 76] Since the above-edge transitions probe the unoccupied density of states above the Fermi level, a reduction of intensity right above the edge reflects the filling of unoccupied states, similar to the band filling found in optical studies of semi-conductors.[77]

In the systems studied here, the VB is dominated by the halogen n*p* orbitals (in the present case Br 4p or Br4p/Cl3p) and to a lesser extent, by the Pb 6s orbitals.[38, 45, 47, 49, 50] The holes created in the VB upon photoexcitation can either be delocalized or become localized at defects or a regular lattice site via a self-trapping process. In the case of delocalization, only a tiny fraction of the charge is "felt" by each atom and no edge shift can be detected. In the case of localization, the removal of a full or partial electron charge, i.e. the localization of a full or partial hole charge, turning the bromide to $Br^0(4p^5)$ in the case of a full charge, would lead to an edge shift to higher energies. In addition, this would open a channel for the 1s-4p transition that lies just below the edge. This behavior of the XAS transients has been reported after electron abstraction of photoexcited aqueous $Br^-$ at the K-edge[68] and aqueous $I^-$ at the $L_1$-edge.[78] It is reproduced here with the resonance at 13.477 keV attributed to the 1s-4p transition, while the minimum at 13.48 keV and the maximum at 13.486 keV are typical of a first derivative-like shape due to a blue shift of the edge.

An estimate of the magnitude of the photo-induced blue shift of the Br K-edge (Figure 1) can be made by taking the difference of the blue-shifted ground state spectrum minus the unshifted spectrum. This approach works quite well, as demonstrated in the case of photo-ionized halides in aqueous solutions,[68, 78] and charge transfer in metal complexes.[60, 79] However, it neglects the other sources of edge shift mentioned above such as structural changes, and it does not account for the changes in the occupancy of orbitals below the edge, which correspond to the pre-edge (bound-bound) transitions. Bearing in mind these limitations, we calculated the difference spectra for shifts of +1 to +8 eV, which are compared to the experimental transient in Figure S13. In making these differences, we used the PFY spectrum (Figure S.8a) because the wide energy range explored allows a fine pre- and post-edge normalization. However, since the transients were recorded in TFY mode (in § S2), this may cause an additional source of deviation between the experimental transients and calculated difference spectra. Although none of the latter really satisfactorily matches the first derivative-like shape between 13.48 and 13.49 keV, the 5-6 eV shift is the best at capturing the 13.48 keV to 13.485 keV region. In the case of aqueous bromide turning to neutral bromine upon electron abstraction, Elles *et al*[68] reported an edge shift of ~5 eV, which is close to the present value. We stress again that the deviations below 13.48 keV (appearance of the 1s-4p resonance around 13.475 keV) cannot be accounted for in the difference spectra. Also, the deviations above 13.485 keV are due to structural changes in the environment of $Br^-$ ions after they have been oxidised. In particular, the Br K-edge has been reported to be very sensitive to and undergo significant edge shifts depending on its environment, even without a change of the oxidation state.[72-74]

The appearance of the 1s-4p resonance at 13.475 keV and the estimated oxidation shift of ~5 eV both point to a full positive charge localizing at the Br centres, by analogy with case of photoexcited aqueous $Br^-$.[68] Charge localization may be caused by small hole polaron formation either at a defect or at a regular site of the lattice. In either case, the consequence of the charge localization is a bond elongation between the Br atoms, that have turned neutral, and their nearest neighbors. We suggest that formation of small polarons at regular sites is the main outcome of the photoexcitation, based on the following: a) The transient exhibits rather clear modulations above the edge. Since defects exist in a distribution of geometries, charge trapping at them would lead to a washing out of the above-edge modulations, as was the case in our previous study on anatase $TiO_2$.[63] The fact that the transient shows a rather well defined modulation above the edge points to the excited charge carriers being localized at well-structured sites, which can only be regular Br-sites of the lattice; b) the high quantum yield of 90% for the band gap PL also points to the overwhelming majority of charge carriers being in the regular lattice, rather than at defects. The localization of a full positive charge at Br atoms of the regular lattice implies formation of a small hole polaron via coupling to phonons. Further studies are planned to confirm this hypothesis on single crystals, along with theoretical simulations of the structural changes around small Br-hole polarons.

The kinetic traces in figure 2 were phenomenologically fit using a bi-exponential function convoluted to the instrument response function of 80 ps, and the time constants and pre-exponential factors are given in table 1. While the long component is almost identical in both cases (100-130 ns), the short one decreases by a factor of 2.5-2.7 from $CsPb(ClBr)_3$ to $CsPbBr_3$. Our purpose here is not to interpret these timescales, which require further studies but we stress the analogy with the RT PL decay kinetics of organic-inorganic perovskites. Indeed, a biexponential decay has typically been reported for the latter,[15, 30, 32, 80, 81] with the short component usually on the order of a few nanoseconds, and the long one spanning from tens to hundreds of ns. The origin of the bi-exponential behavior in the case of organic-inorganic perovskites PL is still debated but the fact that the present samples exhibit a similar kinetic behavior hints to a direct connection between the X-ray transients and the PL.

Turning now to the Pb $L_3$-edge spectrum, it is known to be rather featureless as found in the case of various lead oxides with different oxidation states and our steady-state spectrum, shown in Figure 3, reflects the same aspect.[69, 74] This is due to the very short core-hole lifetimes, which broaden all transitions, making their assignment difficult. Using high energy resolution fluorescence detection (HERFD) X-ray absorption near-edge spectroscopy (XANES), which exploits the apparent reduction in the core-hole lifetime broadening, Glatzel and co-workers[70] could unravel more details of the Pb XANES in various $Pb^{2+}$-containing compounds, in particular PbO, whose Pb $L_3$-edge

XANES and HERFD spectra are reproduced in Figure S11 and compared with the XANES spectrum of our sample. The HERFD spectrum of PbO reveals three features at 13.032 keV, 13.042 keV and 13.054 keV, which are also visible in the TFY XANES spectrum. However, our spectrum does not show such clear-cut features. Furthermore, these are expected to lie at different energies since it is a different compound. The analysis of the spectra in Ref. [74] was carried out by simulating the density of states in order to reproduce the spectra. It was concluded that the pre-edge feature at 13.032 keV is dominated by the Pb p-orbitals, but it arises in the $L_3$-edge spectrum due to their strong mixing with Pb d-orbitals. At higher energies, the spectrum is dominated by the Pb d-orbitals, with a weak contribution of Pb p-orbitals. Interestingly, O-ligand orbitals also contribute to the 13.032 keV and 13.054 keV features in the PbO spectrum.

The $L_3$-edge transients in Figure 3 cannot be interpreted in terms of a chemical edge shift, which would be caused by a change of oxidation state of the $Pb^{2+}$ ions. In the present case, this would turn them to $Pb^+$ centres, and the electron localization would have led to an edge shift to lower energies, which can be ruled out based on Figure S14. Rather, the decreased intensity right at the edge (i.e. above the Fermi level), is due to the fact that upon laser excitation, electrons are transferred to the CB (i.e., above the Fermi level), filling the Pb 6p and 6d orbitals, where they remain delocalized at 100 ps time delay, thus reducing the transition probability as a result of the lower density of unoccupied states. Fully delocalized charges in the CB imply that the oxidation shift of the edge is negligible. Concerning the positive features around 13.040 and 13.065 keV, structure changes may well cause the latter, which is expected if polaron formation occurs around the $Br^0$ atoms. On the other hand, the appearance of the 13.040 keV resonance in the chlorinated sample points to an electronic effect. Since the density of unoccupied states of Pb has decreased after excitation, a positive feature originating from them is excluded and the most likely candidates would be Br-ligand orbitals. In PbO the edge regions have a significant contribution from ligand O orbitals.[70] The same applies here, since it is known that the CB contains a small contribution of halogen orbitals.[38, 47] The pre-edge signals reflect the increased density of unoccupied states of the Br-ligands after photo excitation.

Finally, the transient signal at the Cs $L_2$-edge shows no signature of an electronic (at the edge) or structural (above the edge) change. As far as the former is concerned, this is understandable since density of states (DOS) calculations of organic and inorganic perovskites show that the states of the cation are either far from below the top of the VB or far above the bottom of the CB[38, 45-47] [8, 38, 48-52] and therefore, they would not contribute to the charge carrier dynamics under the present excitation conditions. As for the lack of signature on the above edge region is concerned, we note that already the steady-state ground state spectrum is featureless in this region and is rather insensitive to structure. We conclude that the cation is not a direct player in the fate of the charge carriers, at least for time scales >80 ps.

There is an on-going debate about electron and hole mobilities in organic-inorganic perovskites[16, 17, 21, 82], while only one theoretical study exists for inorganic ones.[83] As far as the former are concerned, Zhu and Podzorov[82] have suggested that charges carriers in organic-inorganic perovskites may be protected by the formation of a large (heavy) polaron, which may also explain the modest charge carrier mobilities. Brenner et al[21] have suggested that the relatively (compared to more conventional semi-conductors) low mobility of charge carriers is due to a strong electron-phonon coupling. They argued that since the mobility is proportional to the carrier scattering time and inversely proportional to the effective mass, and since effective mass values of organic-inorganic perovskites are similar to those of common inorganic semiconductors, then the mobility must be limited by scattering, which can either be due to electron-phonon coupling or impurity scattering. They suggest that this is due to the former, which can also explain the low mobilities and the long lifetimes of the charge carriers. The small hole polaron formation inferred from our results is compatible with a strong electron-phonon scattering. These considerations concern the organic-inorganic perovskites but carrier mobilities appear to be significantly higher in inorganic ones, which hints to different charge transport properties.[7, 84, 85] Using the hybrid density functional theory, Neukirch *et al*[83] found that in a $CsPbI_3$ cluster electron and hole polarons form with larger BE for the electrons. This does not seem to be borne out by the present results. This debate about charge mobilities stresses the need for further experimental and theoretical studies.

## V.     Conclusions

In summary, we have used picosecond X-ray absorption spectroscopy to investigate the fate of the charge carriers in photo excited inorganic pure (Br) and mixed halide (Br-Cl) perovskites nanoparticles in solution. The main conclusions are that at times ≥100 ps, holes are fully localized at Br atoms, forming small polarons, while electrons remain delocalized in the CB. The Cs cation does not show any effect upon photoexcitation leading to the conclusion that it does not play a role in the charge transport and localization, at least at times >80 ps. The decay of the Br K-edge transient signal is bi-exponential, showing the same trends as the optical PL of organic-inorganic compounds, with the long component being of the same order of magnitude.

The conclusion of localized holes and delocalized electrons is reminiscent of the weakly localized exciton in organic-inorganic perovskites discussed by He *et al*,[32] who attributed them to band tail states. In their picture, these band tail states are mainly due to electrons at the bottom of the CB. Our results suggest that on the contrary, it is the hole that is localized rather than the electron. This issue also relates to the debate about the relatively low charge carrier mobilities of perovskites, and the present studies call for more detailed investigations.

On a broader perspective, the present work highlights the advantages of a time-resolved element-selective tool for the study of charge carrier dynamics in multi-element materials. While time-resolved XAS was previously used to study photo excited solids such as $TiO_2$,[63, 64] it mainly detected trapping of charge carriers at defects. In the present work we probed with atomic-selectivity for the first time, the time evolution of the atomic actors involved in the fate of the charge carriers in a multi-element semi-conducting solar material where the role of defects is minimized. Extensions into the femtosecond time domain of such experiments are being planned and will deliver insight about the charge transport and the time scale of localization, which are crucial for improving the performances of such materials.

## Acknowledgments:

This work was supported by the NCCR:Molecular Ultrafast Science and Technology (MUST), the NRP70 "Energy Turnaround" project n° 407040_154056 and projects 200020_155893, 200020_169914 , 200021_135009 and 200021_143638 of the Swiss SNF. Support from the SEFRI via the COST project C13.0062/CM1202 and the European Union via FP7 European Research Council Starting Grant (No. 306733) and European Union's Horizon-2020 program through the Marie-Sklodowska Curie ITN network PHONSI (H2020-MSCA-ITN-642656) are also acknowledged. We thank the Scientific Centre for Optical and Electron Microscopy (ETH Zurich) and the Swiss Light Source for their support and for beam time.

**Table 1: Parameters of the biexponential fit of the kinetic traces of Figure 2.**

| Sample | $\tau_1(A_1)$ | $\tau_2(A_2)$ |
| --- | --- | --- |
| CsPb(ClBr)$_3$ | 195±25 ps (72%) | 132±30 ns (28%) |
| CsPbBr$_3$ | 542±44 ps (64%) | 104±14 ns (36%) |

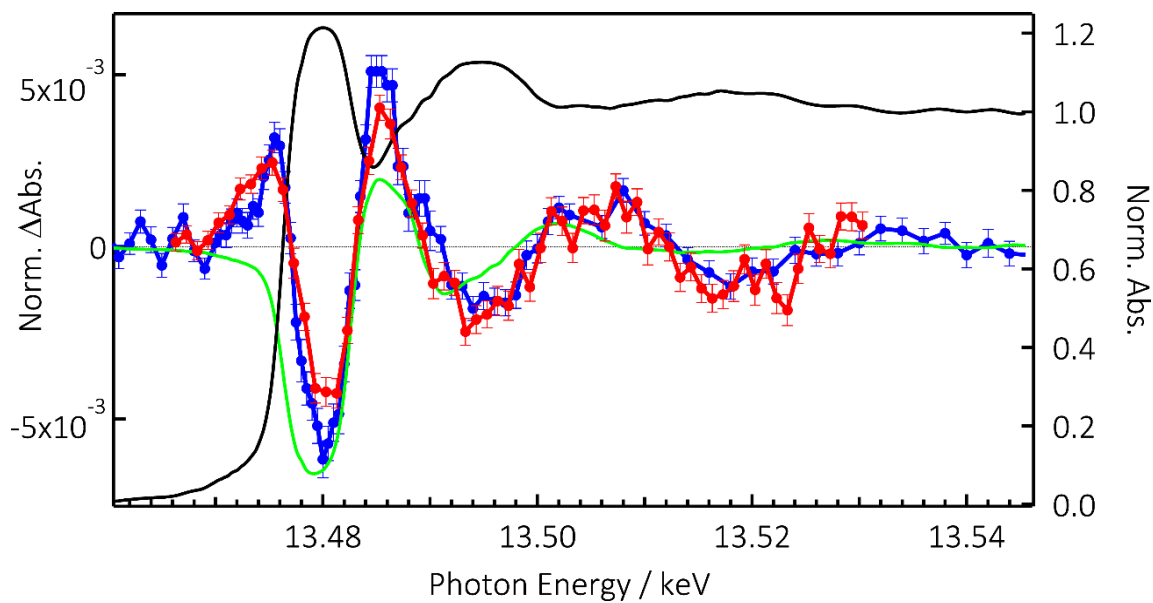

Figure 1: Partial Fluorescence Yield (PFY) X-ray absorption spectrum of CsPb(ClBr)$_3$ at the Br K-edge (black trace) with transients at 100 ps of CsPb(ClBr)$_3$ (blue trace) and CsPbBr$_3$ (red trace) excited at 355 nm with 15 mJ/cm$^2$. The transients were recorded in total fluorescence yield (TFY) detection mode. The green trace represents the difference of the steady state spectrum (black trace) shifted by +5 eV minus the unshifted steady-state spectrum.

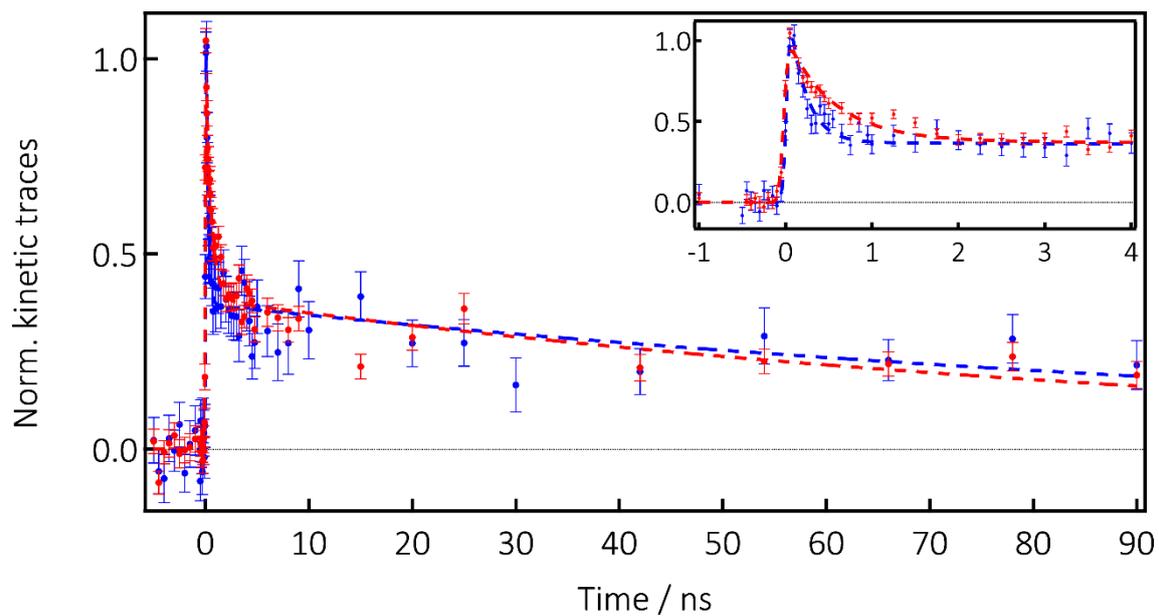

Figure 2: Kinetic traces of the X-ray signal at 13.048 keV upon photoexcitation at 355 nm of CsPb(ClBr)$_3$ (blue dots) and CsPbBr$_3$ (red dots) NCs using a fluence of 15 mJ/cm$^2$. The dashed lines represent a bi-exponential fit (see § S.10) yielding the fit parameters listed in table I. The inset is a zoom of the first 4 ns after excitation.

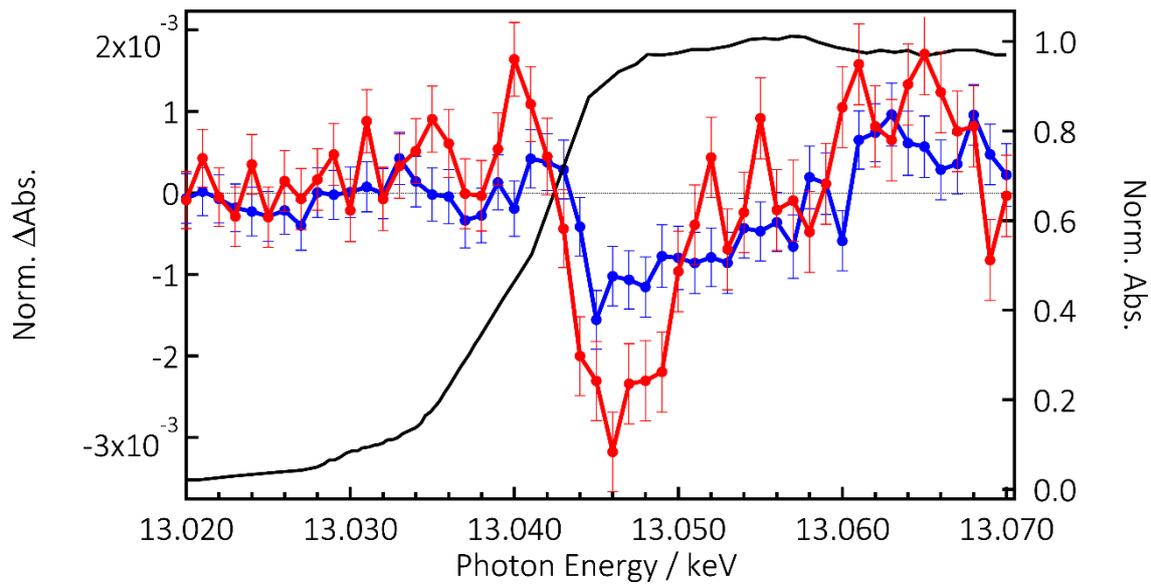

Figure 3: PFY Steady-state X-ray absorption spectrum of CsPb(ClBr)$_3$ NCs at the Pb L$_3$-edge (black trace) and transient spectra at 100 ps time delay after photoexcitation at 355 nm with 15 mJ/cm$^2$ of CsPb(ClBr)$_3$ (blue trace) and CsPbBr$_3$ (red trace).

# Supplementary Information

# Localized holes and delocalized electrons in photoexcited inorganic perovskites: Watching each atomic actor by picosecond X-ray absorption spectroscopy

Fabio G. Santomauro[1], Jakob Grilj[1], Lars Mewes[1], Georgian Nedelcu[2,3], Sergii Yakunin[2,3], Thomas Rossi[1], Gloria Capano[1], André Al Haddad[1], James Budarz[1], Dominik Kinschel[1], Dario S. Ferreira[4], Giacomo Rossi[1,‡], Mario Gutierrez Tovar[1,§], Daniel Grolimund[4], **Valerie** Samson[4], Maarten Nachtegaal[4], Grigory Smolentsev[4], Maksym V. Kovalenko[2,3] and Majed Chergui[1]

[1] Laboratoire de Spectroscopie Ultrarapide, ISIC-FSB and Lausanne Centre for Ultrafast Science (LACUS), Ecole Polytechnique Fédérale de Lausanne, CH-1015 Lausanne, Switzerland

[2] Institute of Inorganic Chemistry, Department of Chemistry and Applied Biosciences, ETH Zürich, CH-8093 Zürich, Switzerland

[3] Laboratory for Thin Films and Photovoltaics, Empa − Swiss Federal Laboratories for Materials Science and Technology, CH-8600 Dübendorf, Switzerland

[4] Paul Scherrer Institut, CH-5232 Villigen, Switzerland

Majed.chergui@epfl.ch

**S1      Synthesis of $CsPbX_3$ NCs and characterization**

The synthetic procedure for the preparation of suspended perovskite nanocrystals (Figure S1) is based on the recipe presented in the work of Protesescu *et al.*[1] with minor changes as discussed below.

**Synthesis of $CsPbBr_3$ NCs:** 11.28 mmol $PbBr_2$ (ABCR, 98%), 200 mL ODE (Aldrich, 90%) were loaded into a 500 mL 3-neck flask and degassed for 45 minutes under vacuum at 120 ºC. 30 mL of dried oleic acid (Sigma-Aldrich, 90%) and 30 mL of dried oleylamine (Acros Organics, 80-90%) were injected at 120 ºC under $N_2$ flow. The temperature was raised to 200 °C and 32 mL of stock solution of Cs-oleate (prepared as described in Ref.[1]) was swiftly injected and 40 seconds later the reaction mixture was cooled down by the water bath.

---

[‡] On leave from : Dipartimento di Fisica e Astronomia, Alma Mater Studiorum - Università di Bologna Viale Berti Pichat 6/2, 40127 Bologna, Italy

[§] On leave from : Departamento de Química Física, Facultad de Ciencias Ambientales y Bioquímica, y INAMOL, Campus Tecnolóogico de Toledo, Universidad de Castilla La Mancha (UCLM), Avenida Carlos III, S.N., 45071 Toledo, Spain.

**Synthesis of CsPb(ClBr)$_3$ NCs:** 0.392 mmol PbCl$_2$ (ABCR, 99.999%), 0.730 mmol PbBr$_2$ (ABCR, 98%), 20 mL ODE (Aldrich, 90%) and 4 mL TOP (Strem, 97%) were loaded into a 100 mL 3-neck flask and degassed for 45 minutes under vacuum at 120 °C. 3 mL of dried oleic acid (Sigma-Aldrich, 90%) and 3 mL of dried oleylamine (Acros Organics, 80-90%) were injected at 120 °C under N$_2$ flow. The temperature was raised to 200 °C and 3.2 mL of stock solution of Cs-oleate was swiftly injected and 40 seconds later the reaction mixture was cooled down by the water bath.

**Isolation and purification:** The formed product was isolated by centrifugation and the supernatant was discarded. Hexane (1.2 mL, Sigma-Aldrich, ≥ 95%) was added to disperse the NCs and centrifuged again. After centrifugation, the supernatant was discarded and 0.5 mL of hexane was added to the precipitate and centrifuged again. The resulting precipitate was re-dispersed in 7 mL toluene (Fischer Scientific, HPLC grade). Finally, the samples were diluted using a solution of toluene with 0.2 % of 1:1 mixture of oleylamine and oleic acid until 1.2 of optical density (OD) at 355 nm for X-ray.

**Transmission electron microscopy (TEM)** images were recorded using Philips CM 12 microscope operated at 120kV. Figure S2 shows the images for the two samples, which are both monodispersed with an average size of 12 nm.

**Powder X-ray diffraction patterns (XRD)** were collected with STOE STADI P powder diffractometer, operating in transmission mode. Germanium monochromator, Cu Kα1 irradiation and silicon strip detector Dectris Mythen were used. Figure S3 and S4 show the diffraction patterns for the two samples, compared with reference patterns of CsPbBr$_3$ and CsPbCl$_3$.[2] The diffraction peaks of CsPb(ClBr)$_3$ are between the one of these two references which indicates the hybrid nature of this sample. Indeed, as indicated by the amount of precursors in the synthesis, it should result in a final stoichiometry of 2 Br$^-$ per each Cl$^-$.

**S2. Experimental procedures**

The X-ray experiments were performed at the MicroXAS and SuperXAS beamlines of the Swiss Light Source at the Paul Scherrer Institute (Villigen). In both beamlines a Si (111) monochromator was used with a nominal resolution of about 1.5 eV at 13 keV. The steady-state spectra were performed exclusively at microXAS using a Si KETEK detector in partial fluorescence yield (PFY) detection mode by selectively recording the Br K$_\alpha$ and Pb L$_\alpha$ emissions. The ps-XAS measurements were performed using the high repetition rate data acquisition scheme.[3] Transient XAS spectra (X-ray absorbance difference of the excited minus the unexcited sample) were recorded on pulse-to-pulse basis and the transient signals intensity was divided by the edge-jump magnitude. The signal was acquired in transmission with a Si diode detector and in total fluorescence yield (TFY) mode to enhance the signal using two Si avalanche photodiodes perpendicular to the incident X-ray beam. The samples were excited using 355 nm/10 ps pulses with a fluence of 15 mJ/cm$^2$ and at a repetition rate of 260 kHz. The X-ray pulses from the Swiss Light Source have a pulse width of 80 ps, which determines the temporal resolution of the experiments.

The samples consisted of perovskite nanoparticles dissolved in toluene. The perovskite nanoparticles were characterized by transmission electron microscopy, X-ray powder diffraction, UV-visible spectroscopy and XAS, as discussed in § S.1 and shown in figures S2 to S6. The solution was recirculated in a free flowing jet of 200 μm thickness and ~3 mm width, which was tilted by 45° with respect to the incident x-ray beam. The concentration of $CsPbBr_3$ (about 40 mM, estimated by drying and weighing a known volume of suspension) and $CsPb(ClBr)_3$ (about 75 mM) in toluene gives an optical density of 1.2 at 355 nm for both materials.

**S.3 Sample stability under laser irradiation**

The fluence dependence of the X-ray transient signal was recorded (Figure S7) and is linear up to 15 mJ/cm$^2$. Therefore, all the measurements in this work were performed at this latter fluence. After >12 hours of ps-XAS measurements of $CsPb(ClBr)_3$ at 15 mJ/cm$^2$ minimal changes are observed in the absorption and emission spectra (Figure S5), which may simply be due to a change in the scattering properties of the samples. Indeed, this change does not affect the XAS spectra (Figure S6).

**S.4 Pb L$_3$-edge absorption spectra**

Figure S11 shows the Pb L$_3$-edge XANES and EXAFS spectra of PbO from ref.[4], recorded in TFY mode and using the High Energy Resolution Fluorescence Detection (HERFD) mode. The TFY spectrum shows only weak features as is often the case with Pb L$_3$ spectra,[5] while the HERFD spectrum clearly resolves the weak structures at 13.031 keV, 13.040 keV and 13.054 keV. The DOS calculations[4] show that the first of these bands is due to Pb p-orbitals that are mixed with Pb d-orbitals, with an additional contribution from Oxygen p- and s-orbitals. The latter two are mostly due to Pb d-orbitals.

The TFY XANES spectrum of the present $CsPb(ClBr)_3$ sample is also shown in figure S11. It shows weak shoulders at 13.031 keV, 13.040 keV, 13.046 keV and 13.058 keV. In the absence of calculations, it is difficult to propose an assignment, but these features may well have some correspondence with the PbO bands, though and not unexpectedly they are shifted in energy due to a different crystal field.

# Figures

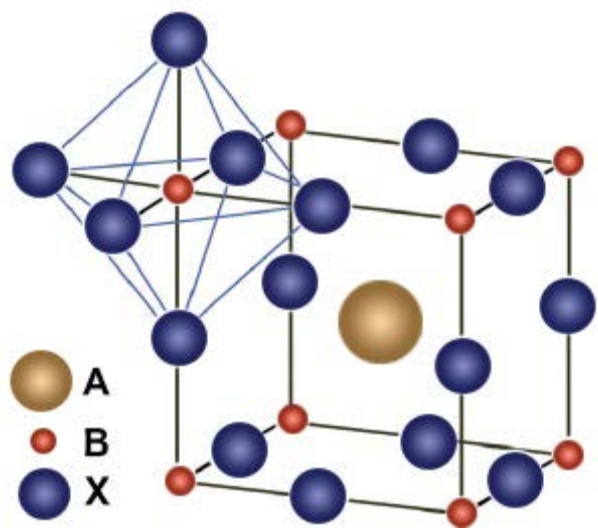

Figure S1: ABX$_3$ perovskite structure where in the present work, A represent the Cs$^+$, B the Pb$^{2+}$ and X the Br$^-$/Cl$^-$ ions.

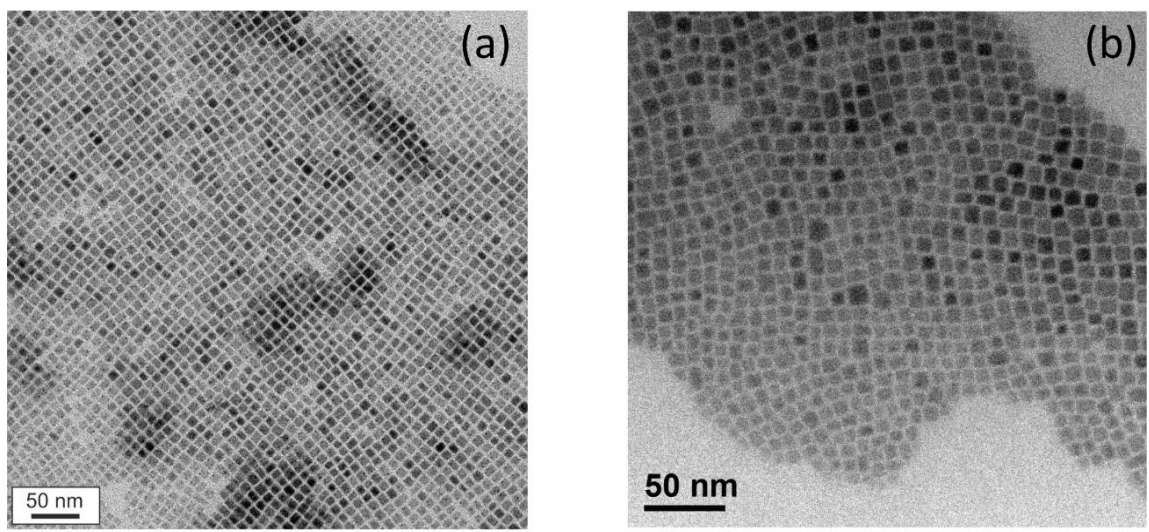

Figure S2: Transmission electron micrograph of monodisperse CsPbBr$_3$ (a) and CsPb(ClBr)$_3$ (b) nanocrystals with an average size of 12 nm.

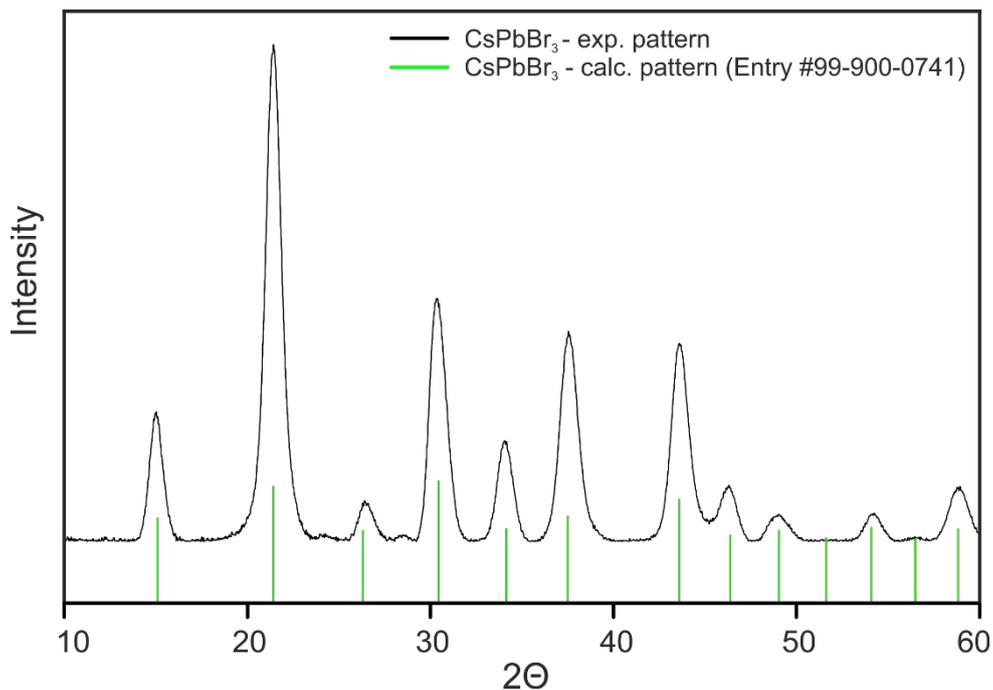

Figure S3: Powder X-ray diffraction pattern of cubic $CsPbBr_3$ NCs matching the cubic phase of the reference pattern for $CsPbBr_3$ (green sticks, space group Pm-3m)[2].

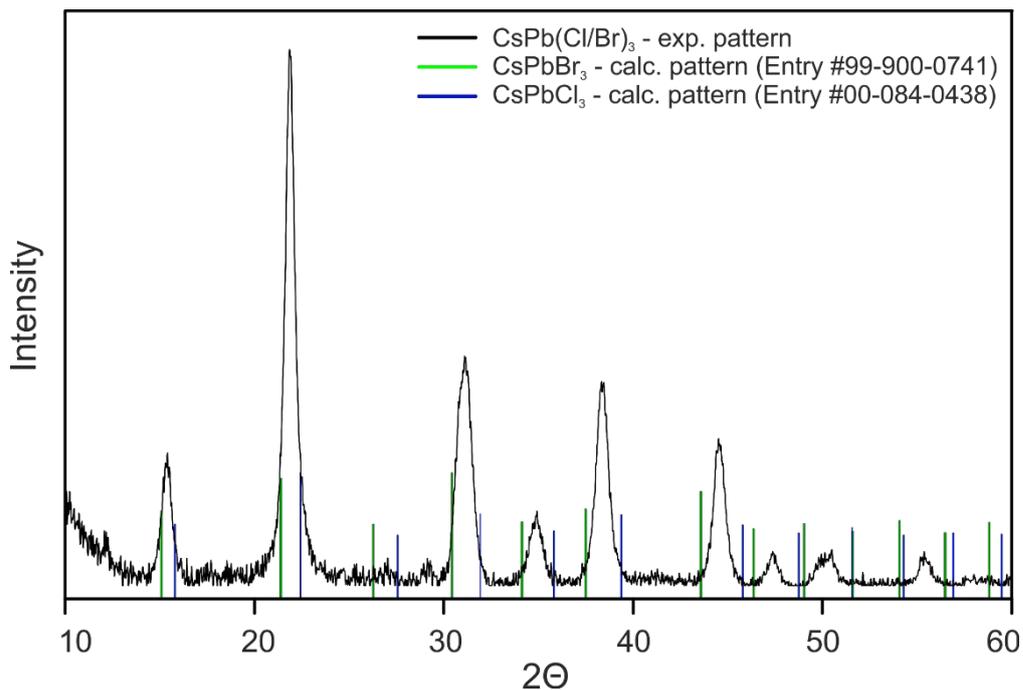

Figure S4: Powder X-ray diffraction pattern of $CsPb(ClBr)_3$ NCs displaying the diffraction peaks in between the reference patterns of $CsPbBr_3$ (green sticks)[2] and $CsPbCl_3$ (blue sticks)[2]. The experimental pattern of $CsPb(ClBr)_3$ presents a homogeneous distribution of the halide anions into the perovskite crystal structure by exposing the diffraction peaks into between the typical XRD patterns of $CsPbBr_3$ and $CsPbCl_3$.

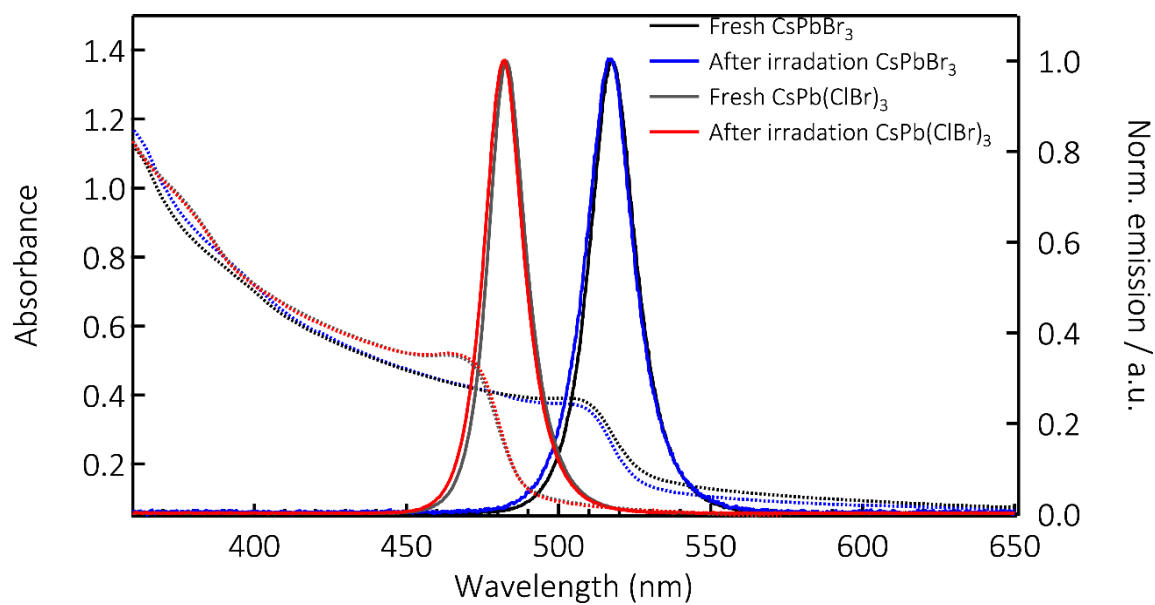

Figure S5: Absorption spectra (dotted) and Photoluminescence spectra (full lines) of CsPb(ClBr)$_3$ (grey and red traces) and CsPbBr$_3$ (black and blue traces) NCs. We also show the same spectra after after >12 hours of laser irradiation at 355 nm, 260 kHz rep. rate and a fluence of 15 mJ/cm$^2$.

.

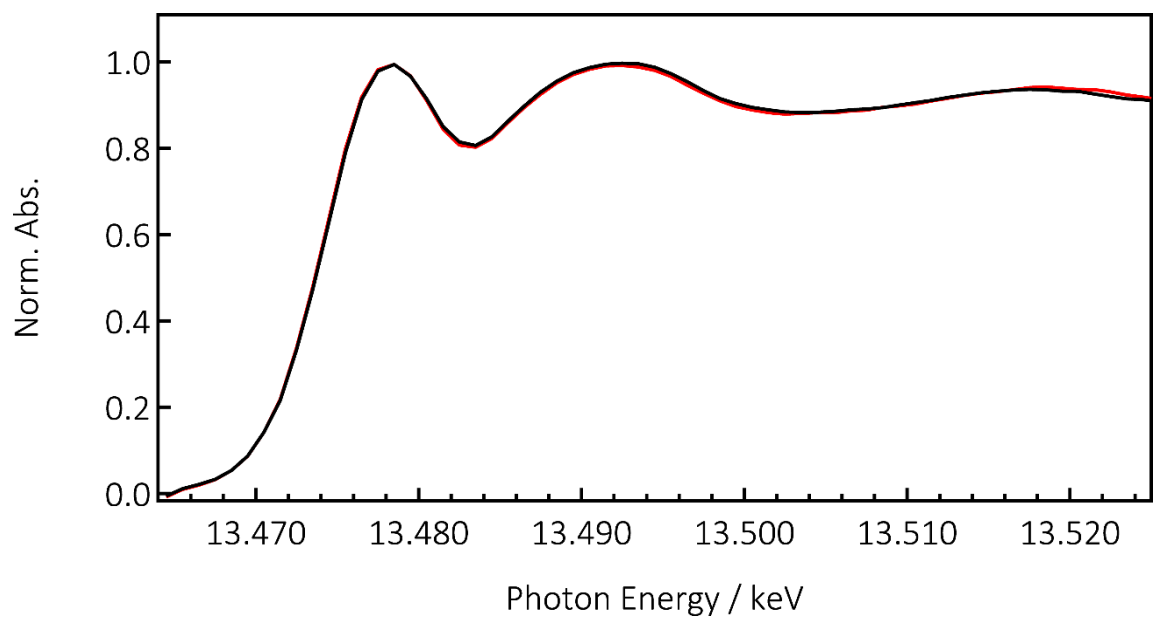

Figure S6: Normalized X-ray absorption spectra of CsPbBr$_3$ near the Br K-edge region acquired in TFY. The agreement between the spectrum averaged over the first 2 hours of data acquisition of the pump-probe experiment (red curve) and that over the consecutive 2 hours (back curve), supports the absence of damage during the measurements upon irradiation at 15 μJ/cm$^2$.

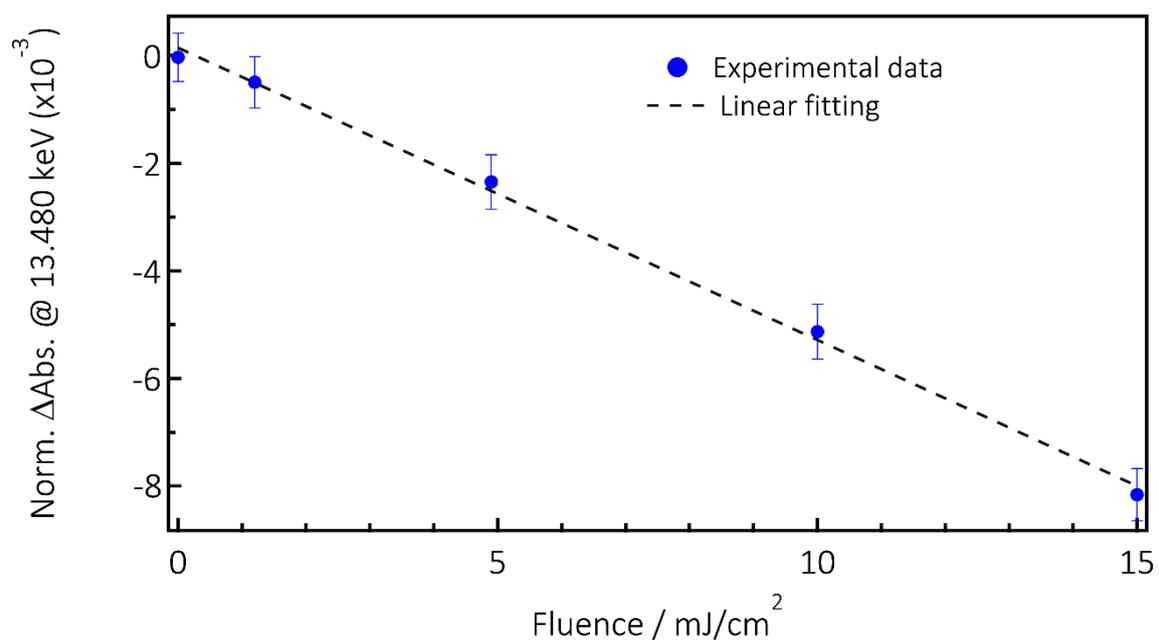

Figure S7: Fluence dependence of the intensity of the Br K-edge transient signal at 13.480 keV for CsPb(ClBr)$_3$ NCs photoexcited at 355 nm. The black dashed line shows a linear fitting of the experimental data (blue dots).

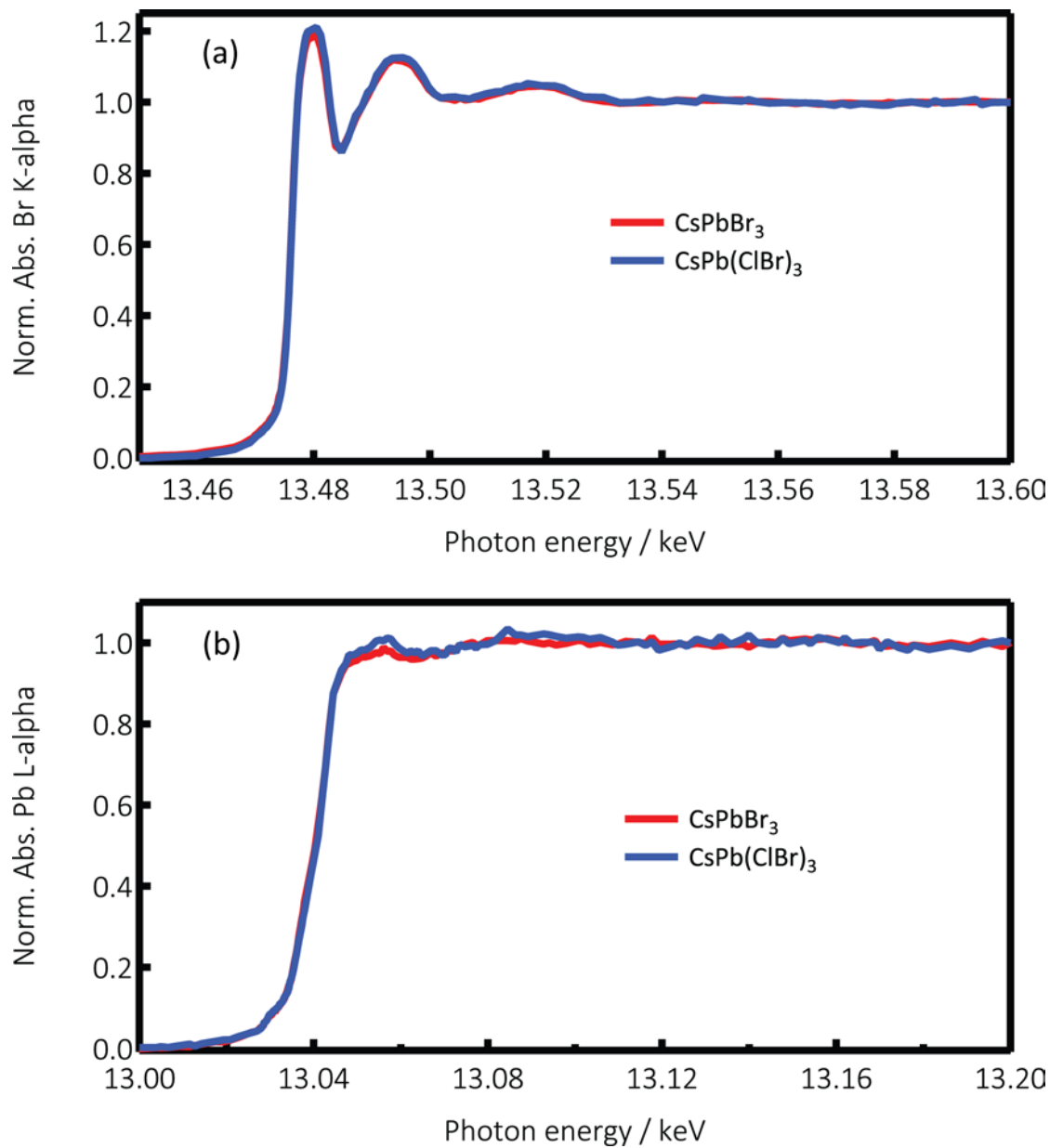

Figure S8: PFY X-ray absorption spectra near the edge region of CsPbBr$_3$ (red trace) and CsPb(ClBr)$_3$ (blue trace) NCs at the Br K-edge (a) and Pb L$_3$-edge (b). The difference in intensity of the Pb L$_3$-edges at about 13.05 keV between the two compounds is related to a higher density of unoccupied states due to the presence of Cl.

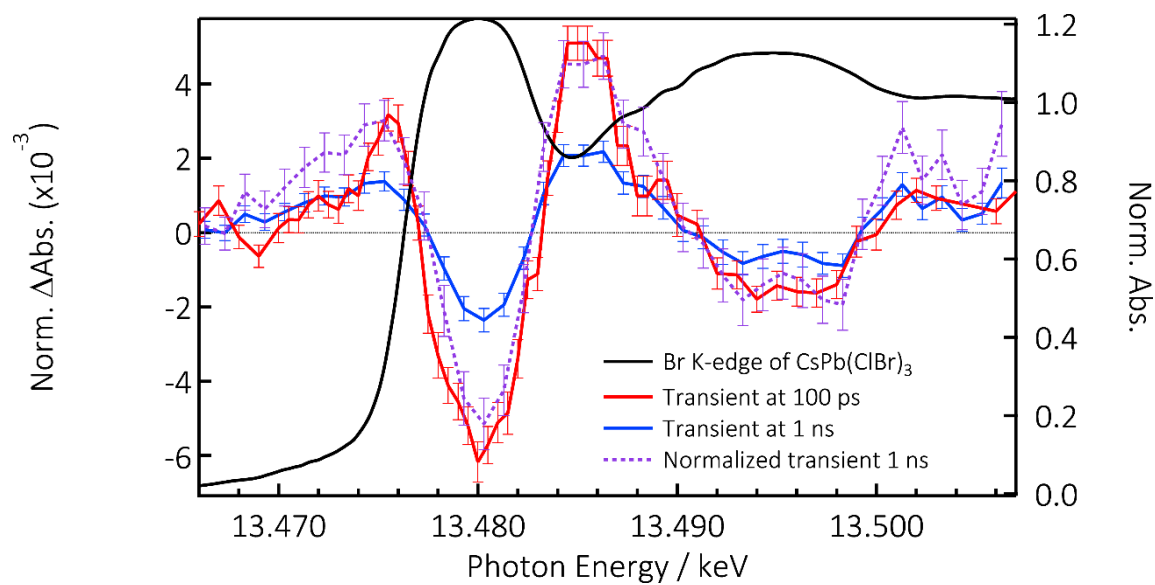

Figure S9: X-ray absorption spectrum of CsPb(ClBr)$_3$ NCs at the Br K-edge (black trace) with the transient spectra at 100 ps (red trace), 1 ns (blue trace) and 1 ns normalized to the 100 ps intensity (violet dotted trace).

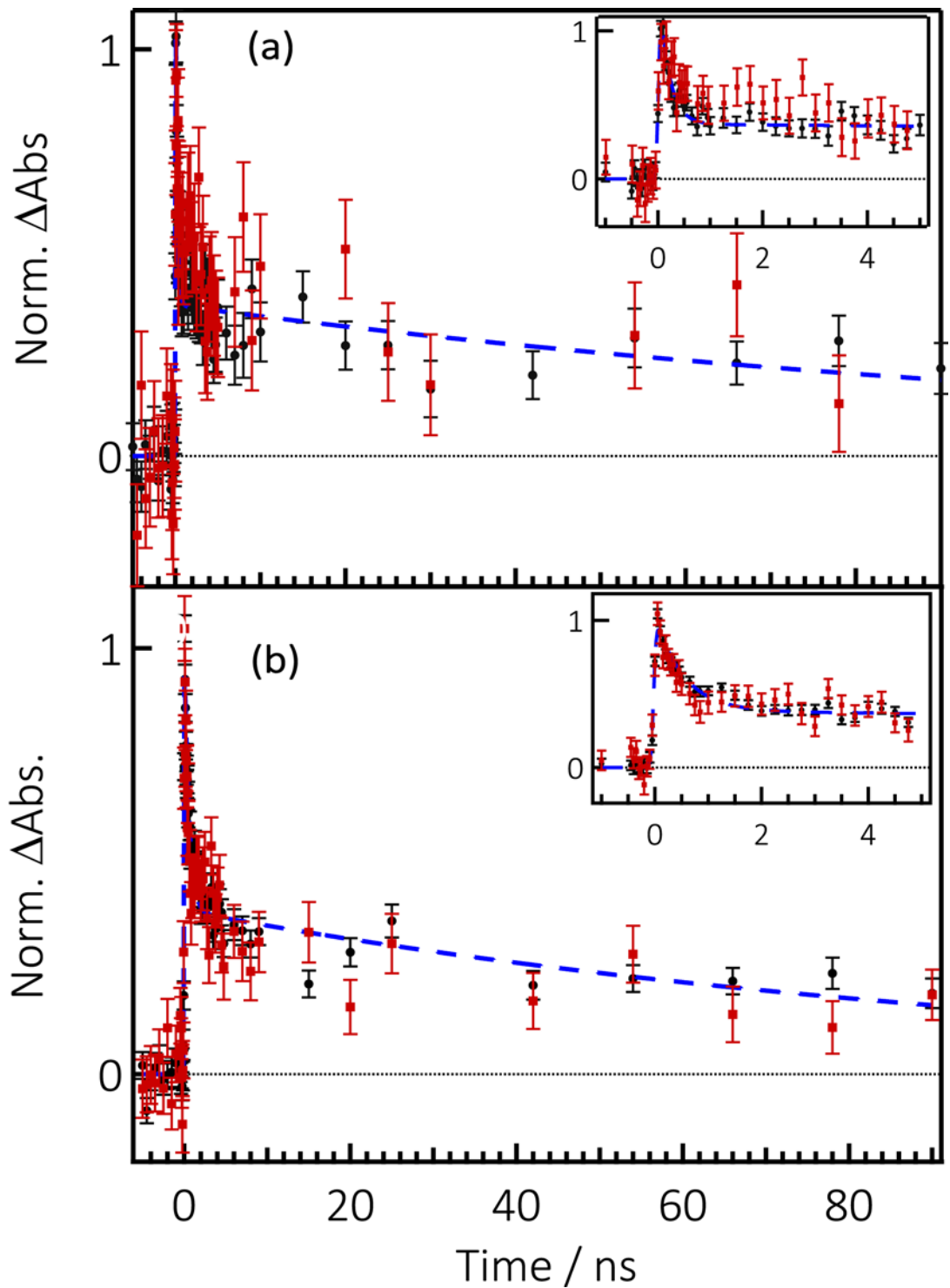

Figure S10: Normalized kinetic traces at 13.480 (black dots) and 13.4863 keV (red squares) and fit using a bi-exponential (blue dotted curve) of CsPb(ClBr)$_3$ (a) and CsPbBr$_3$ (b) NCs upon photoexcitation at 355 nm.

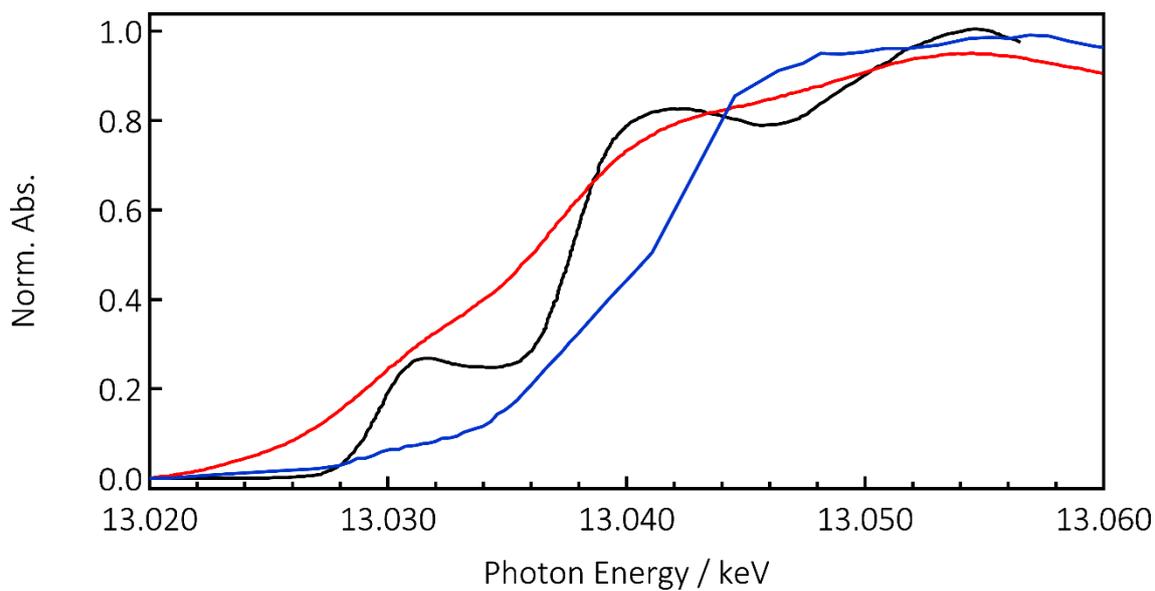

Figure S11: Pb L$_3$-edge spectra of our CsPb(ClBr)$_3$ sample (blue). The XANES (red) and the HREFD (black) for PbO are reproduced from ref. [4].

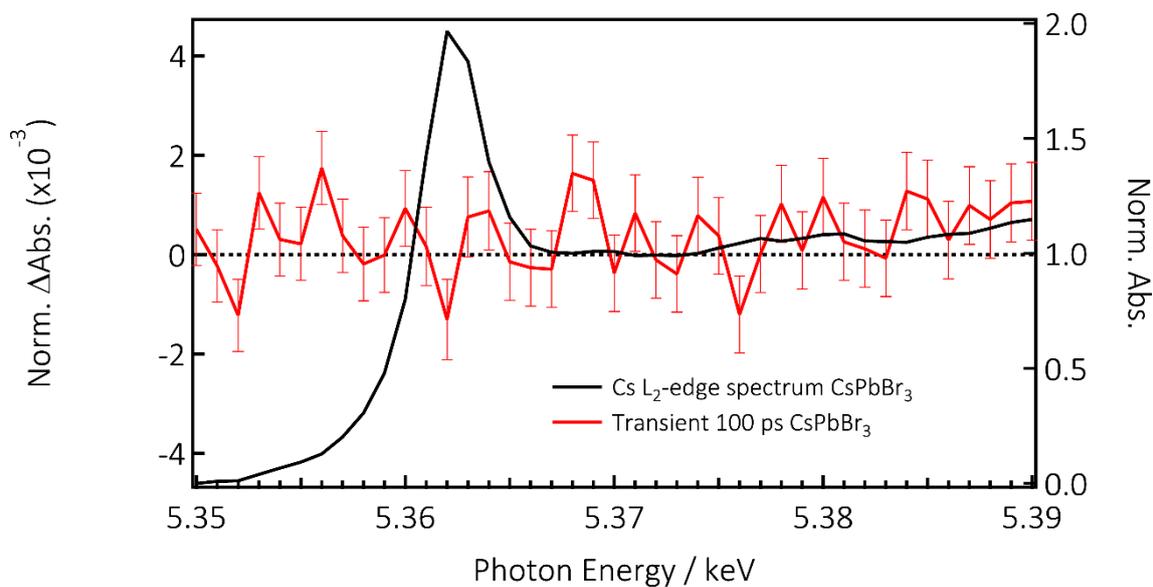

Figure S12: X-ray absorption spectrum of CsPbBr$_3$ NCs at the Cs L$_2$-edge (black trace) with the transient spectra at 100 ps (red trace) showing no signal within the S/N of the measurement.

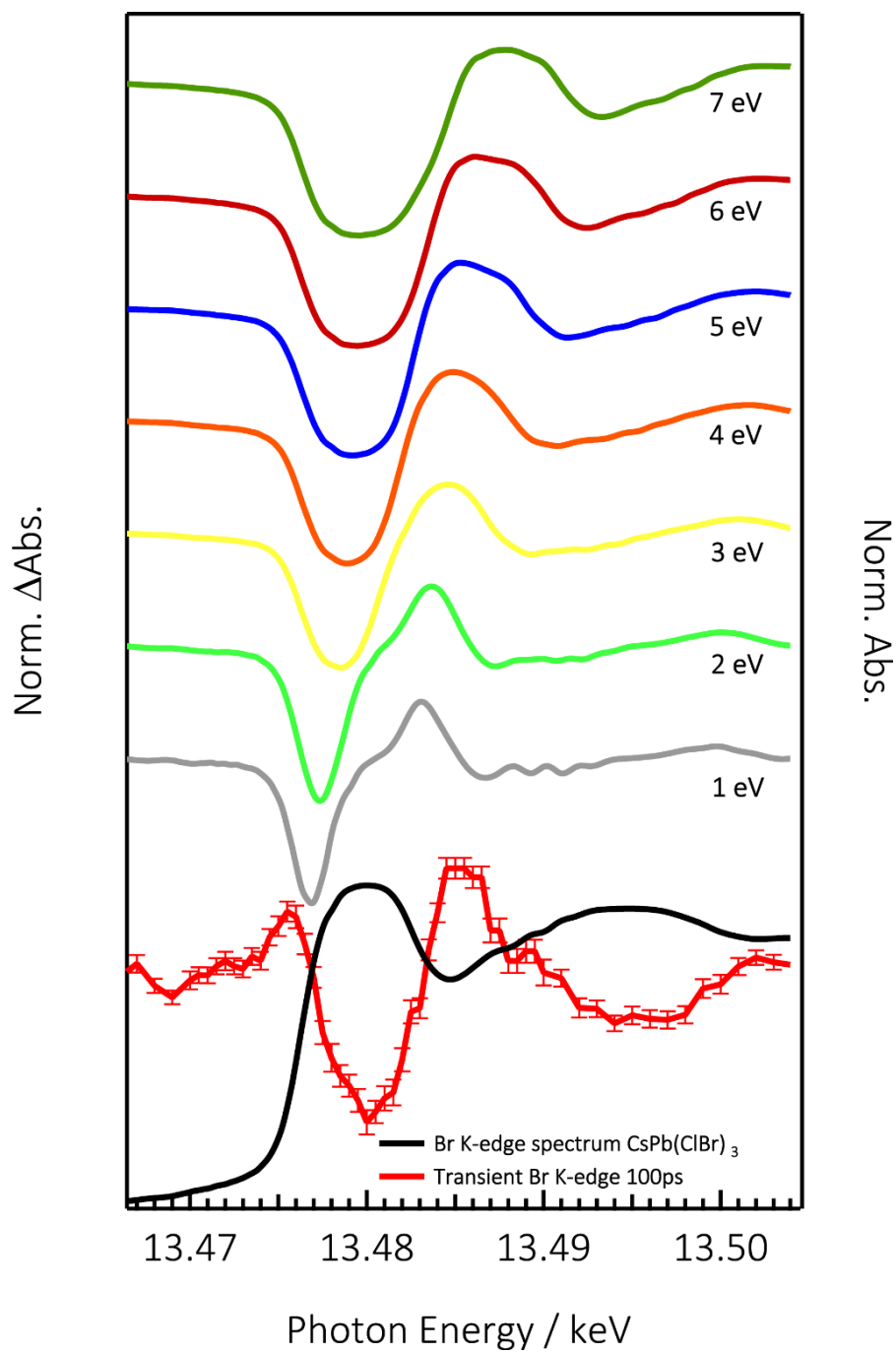

Figure S13: PFY X-ray absorption spectrum of CsPb(ClBr)$_3$ NCs at the Br K-edge (black trace) with the transient at 100 ps upon photoexcitation at 355 nm (red trace) and calculated difference spectra of the ground state spectrum minus the unshifted one for different values of the shift (1 to 8 eV, different offsets on the graph). The difference spectra have been rescaled with factors between 0.05 and 0.006 in order to best match the intensity of the experimental negative transient feature at 13.480 keV.

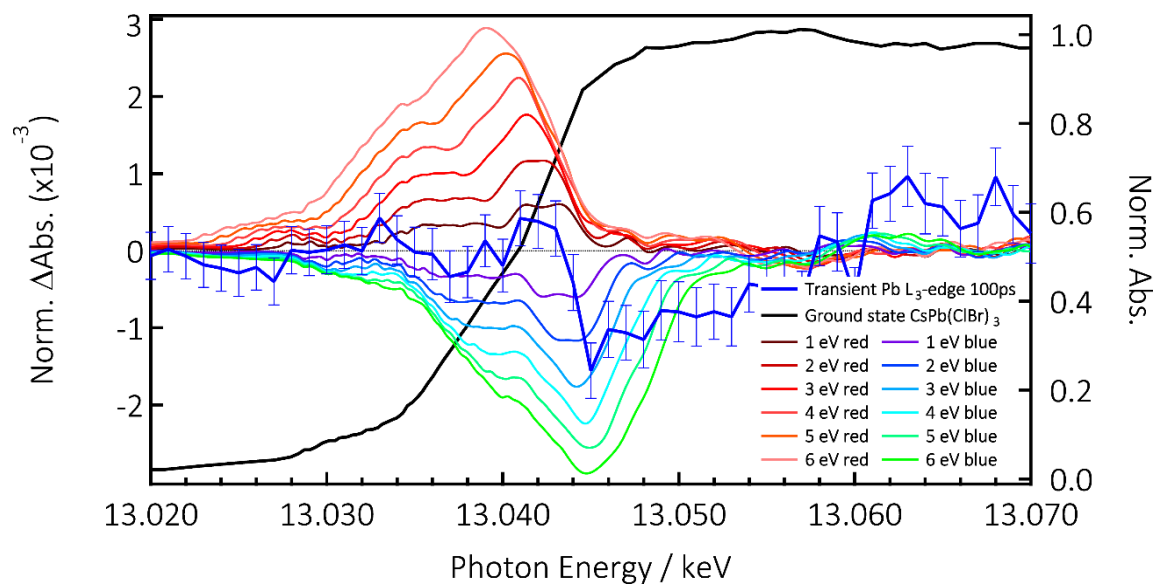

Figure S14: X-ray absorption spectrum of CsPb(ClBr)$_3$ NCs at the Pb L$_3$-edge (black trace) with the transient at 100 ps upon photoexcitation at 355 nm (blue trace) and simulated transient spectra via the rigid shift of different magnitude and signs of the ground state spectrum.